\documentclass[twocolumn,twoside,slac_two]{revtex4}
\usepackage{graphicx}
\usepackage{fancyhdr}
\def \agile {AGILE}
\def \egret {EGRET}

\def \igr {INTEGRAL}
\def \swi {{\it Swift}}

\def \rxte {RXTE}

\def \ergcmsec{\hbox{erg cm$^{-2}$ s$^{-1}$}}
\def \phcmsec{\hbox{photons cm$^{-2}$ s$^{-1}$}}

\def \gray {$\gamma$-ray}
\def \source {3C~454.3}

\def\aap {Astron. \& Astrophys.\/}

\def\apj {Astrophys. J.\/}
\def\apjl {Astrophys. J. Letters}

\def\cjaa {Chinese J. of Astron. \& Astrophys.\/} 
\def\iaucirc {IAU Circ.\/} 
\def\mnras {Mon. Not. R. Astron. Soc.\/}

\begin{document}

\title{A success story: \source{} in the gamma-ray energy band}

%

\author{S. Vercellone}
\affiliation{INAF--IASF Palermo, Via Ugo La Malfa 153,  90146 Palermo, Italy}
\author{on behalf of the AGILE Team}

\begin{abstract}
Since 2007, the blazar \source{} has become the most active and the brightest \gray{} source of the sky, 
deserving the nickname of {\it Crazy Diamond}. The short-term variability in the \gray{} energy band and 
the extremely high peak fluxes reached during intense flaring episodes make \source{} one of the best targets 
to investigate the blazar jet properties. We review almost four years of observational properties of this remarkable 
source, discussing both short- and long-term multi-wavelength campaigns, with particular emphasis on the recent 
flaring episode which occurred on 2010 November 20, when \source{} reached on a daily time-scale a \gray{} 
flux ($E>100$\,MeV) higher than $6.5\times10^{-5}$\phcmsec, about six times the flux of the brightest \gray{} 
steady source, the Vela Pulsar.
\end{abstract}

\maketitle

\thispagestyle{fancy}


	\section{INTRODUCTION}\label{sec:intro}
Among the FSRQs detected at energies above 100~MeV, \source{} (PKS~2251$+$158; $z=0.859$) is certainly one of the most active
at high energy. In the \egret{} era, it was detected in 1992 during an intense \gray{} flaring episode 
\citep{Hartman1992:3C454iauc, Hartman1993:3C454_EGRET} when its flux $F_{\rm E>100MeV}$ was observed to vary within the range
$(0.4-1.4) \times 10^{-6}$\,photons\,cm$^{-2}$\,s$^{-1}$. In 1995, a 2-week campaign detected a \gray{} flux $< 1/5$ of its historical maximum
\citep{Aller1997:3C454_EGRET}.

In 2005, \source{} underwent a  major flaring activity in almost all energy bands \cite{Giommi2006:3C454_Swift}.
In the optical, it reached $R=12.0$\,mag \citep{vil06} and it was detected by \igr{} at a flux\footnote{Assuming
a Crab-like spectrum.} level of $\sim 3 \times 10^{-2}$\,photons\,cm$^{-2}$\,s$^{-1}$ in the 3--200~keV 
energy band \citep{Pian2006:3C454_Integral}.  Since the detection of the exceptional 2005 outburst, several monitoring 
campaigns were carried out to follow the source multifrequency behavior \citep{vil06,vil07,rai07,rai08a,rai08b}. 
During the last of these campaigns, \source{} underwent a new optical brightening in mid July 2007, which triggered 
observations at all frequencies.

During 2007 -- 2010, \agile{} detected and investigated several \gray{} flares
\citep{Vercellone2008:3C454_ApJ,Vercellone2008:3c454:ApJ_P1,Donnarumma2009ApJ_P2,Vercellone2010ApJ_P3, Pacciani2010ApJ_3c454, Striani2010ApJ_3c454,Vercellone2011:ApJL_3C454_nov2010}. 
These observations allowed us to establish a possible correlation between the \gray{} 
(0.1 --10 GeV) and the optical (R band) flux variations with no time delay, or with a
lag of the former with respect to the latter of about half a day.
Moreover, the detailed physical modeling of the spectral energy distributions (SEDs) 
when \source{} was at different flux levels provided an interpretation of the emission mechanism responsible 
for the radiation emitted in the \gray{} energy band, assumed to be inverse Compton scattering of photons from 
the broad line region (BLR) clouds off the relativistic electrons in the jet, with bulk Lorentz 
factor $\Gamma \sim 20$.

In this Paper, we review the main results of both long- and short-term observations,  with particular emphasis on the recent 
flaring episode which occurred on 2010 November 20.

	\section{LONG-TERM MONITORING}\label{sec:long}

\agile{} detected \source{} since the very beginning of its operation, during the Science Verification Phase.
\begin{figure*}[!ht]
\centering
\includegraphics[width=135mm,angle=-90]{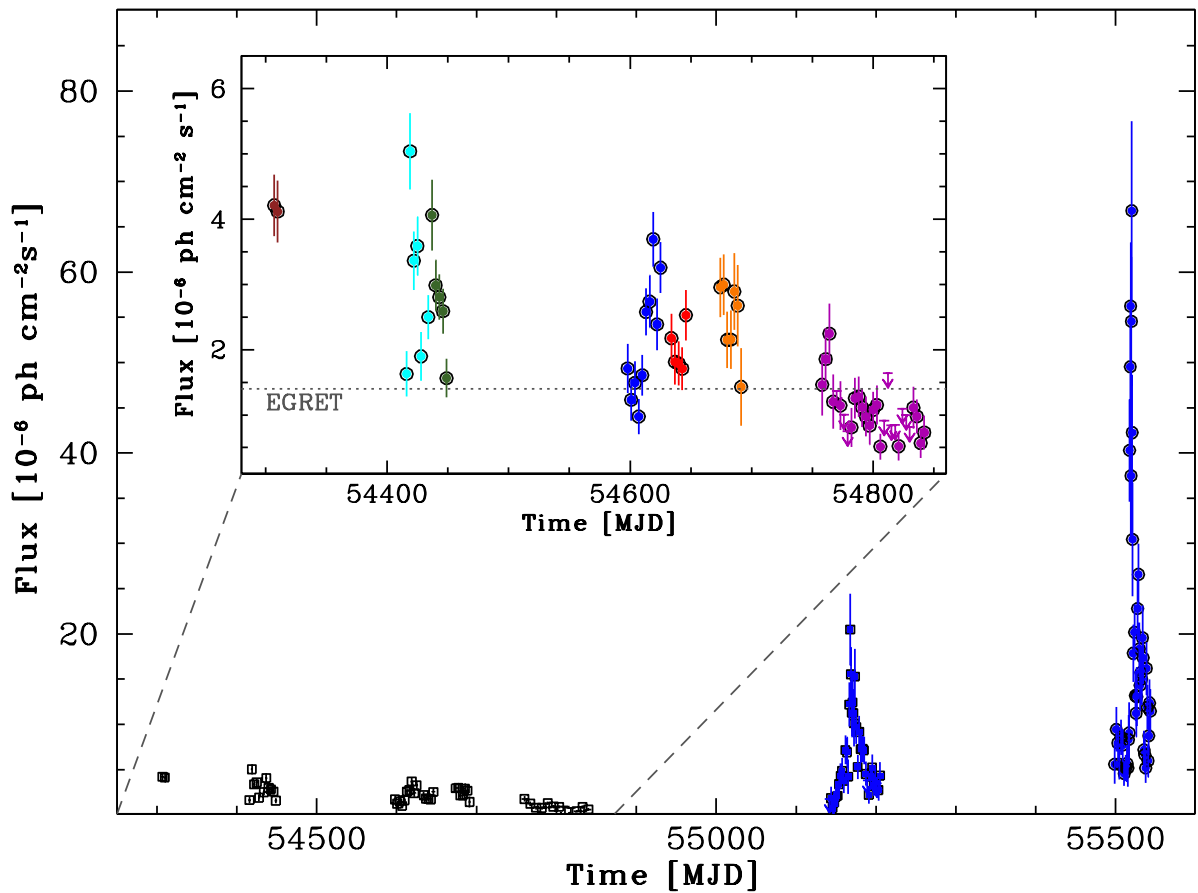}
\caption{
Historical (2007 - 2008, black symbols and inset; data from
\cite{Vercellone2008:3C454_ApJ,Vercellone2008:3c454:ApJ_P1,Donnarumma2009ApJ_P2,Vercellone2010ApJ_P3}) 
and super-flares (2009 - 2010, blue symbols; data from 
\cite{Striani2010ApJ_3c454} and \cite{Vercellone2011:ApJL_3C454_nov2010}, respectively) AGILE \gray{} light-curves
for $E>100$\,MeV in units of $10^{-6}$\,\phcmsec{}.
During the 2007 - 2008 campaign the \gray{} flux level was for most of the time higher than the maximum flux detected by 
EGRET (dotted gray line).
} \label{fig:3c454_monitoring}
\end{figure*}
Figure~\ref{fig:3c454_monitoring} shows the \gray{} light-curve accumulated during the  period
2007 July 15 -- 2010 December 15, with particular emphasis on the \gray{} super-flares which occurred on 
2009 December and 2010 November (blue points), when \source{} reached a \gray{} flux of
$F_{\gamma}^{\rm 2009} = (2.0\pm0.4) \times 10^{-5}$\,\phcmsec{} \cite{Striani2010ApJ_3c454} and 
$F_{\gamma}^{\rm 2010} = (6.8\pm1.0) \times 10^{-5}$\,\phcmsec{} \cite{Vercellone2011:ApJL_3C454_nov2010}, respectively.
We clearly note that the dynamic range in the \gray{} flux is of the order of a factor on 100, assuming as
a low state the flux value during the Fall/Winter 2008 and as the maximum flux level the super-flare
on 2010 November 20.
We note that during the 2007 - 2008 campaign (see the inset for a more detailed view) the \gray{} flux level was, 
for most of the time, higher than the maximum flux detected by EGRET (dotted gray line).
The fast \agile{} data analysis system \cite{Bulgarelli2009:ASPC_DataAnalysis} allowed us to perform several 
multi-wavelength campaigns almost simultaneous with respect to the \gray{} flares, involving both space- 
(e.g., \swi{}, \igr{}, \rxte{}, {\it Spitzer}) and ground-based observatories (GASP--WEBT).

\begin{figure}[!ht]
\centering
\includegraphics[width=65mm,angle=-90]{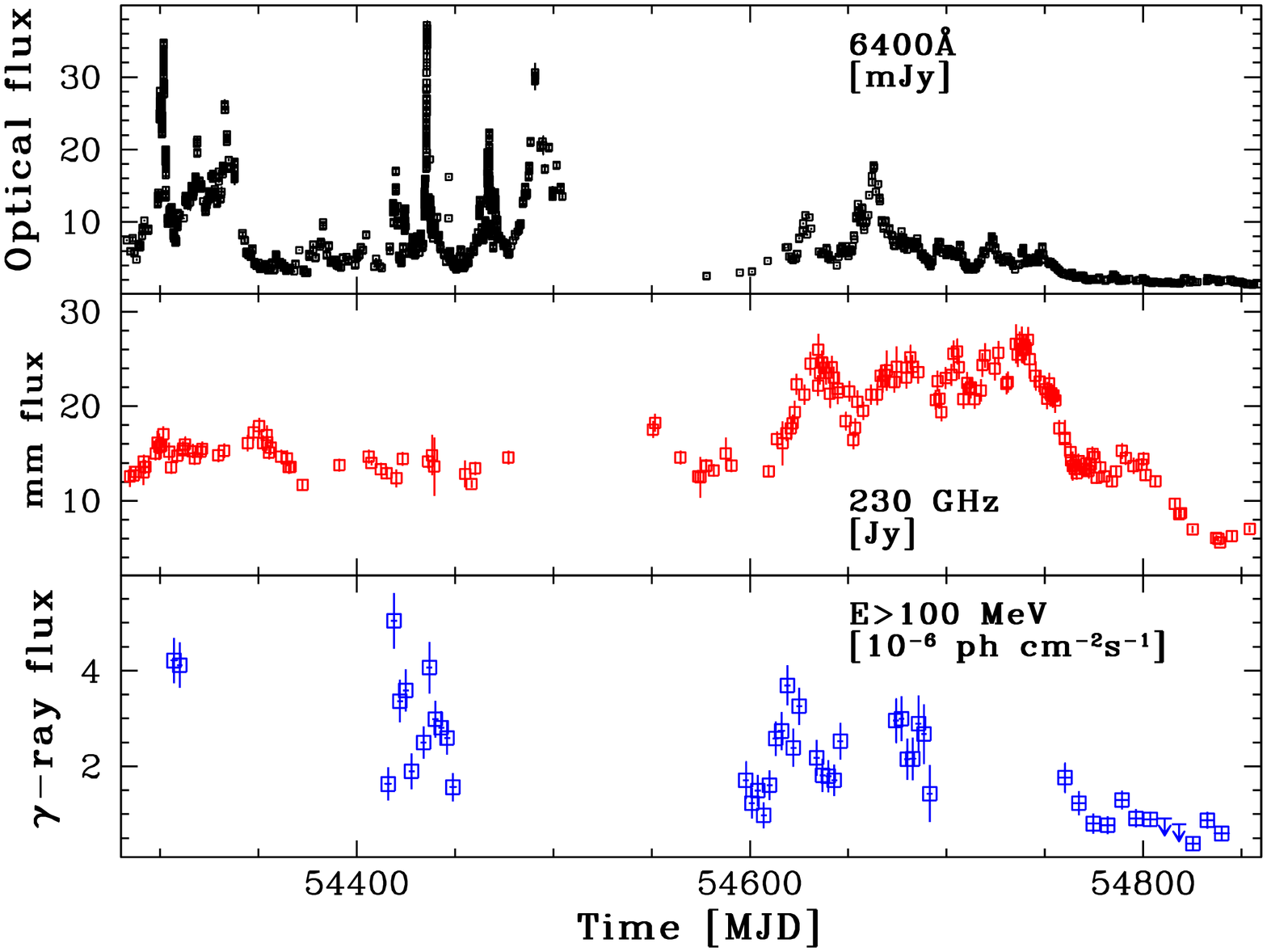}
\caption{
$R$-band, 230~GHz, and \gray{} light-curves (top, middle and bottom panel,
respectively) covering the period 2007 July -- 2009 January. The AGILE light-curve has a time-bin of 1-week.
Adapted from \cite{Vercellone2010ApJ_P3}.
} \label{fig:jet_properties}
\end{figure}

\begin{figure}[!ht]
\centering
\includegraphics[width=80mm,angle=0]{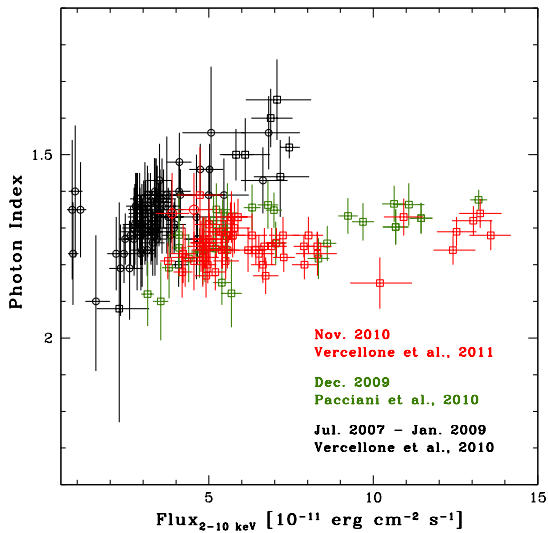}
\caption{
\swi/XRT photon index as a function of the 2--10\,keV flux. Open circles
and squares refer to photon counting and windowed data, respectively.
Black, green and red symbols represent data accumulated during the 2007--2009 monitoring
campaign \cite{Vercellone2010ApJ_P3}, the 2009 December super-flare \cite{Pacciani2010ApJ_3c454}, and
the 2010 November super-flare \cite{Vercellone2011:ApJL_3C454_nov2010}, respectively.
} \label{fig:Xray_hysteresis}
\end{figure}

Figure~\ref{fig:jet_properties} shows the 18-months coverage in the optical ($R$-band), millimeter (230~GHz), and
\gray{} ($E>100$\,MeV) energy bands.
The different variability behavior  at different wavelengths can support the hypothesis of a change in orientation
of a curved jet, yielding different alignment configurations within the jet itself.
In particular, during 2007 the inner portion of
the jet seems to be the more beamed one, because of the co-ordinated optical and \gray{} variability.
On the contrary, during 2008, the more extended region of the jet seems to be more
aligned with respect to the line of sight, as suggested by the enhanced millimeter variability and
by the optical and \gray{} dimming trend. Recently, a detailed model supporting this interpretation
has been presented in \cite{rai11}.
Moreover, the long-term optical and \gray{} coverage allowed us to investigate possible time-lags
between the two energy bands. We obtain (see \cite{Vercellone2008:3c454:ApJ_P1,Donnarumma2009ApJ_P2,Vercellone2010ApJ_P3})
that the emission in the optical band appears to be (weakly) correlated with that at \gray{} energies above 100\,MeV, with a lag
(if present) of the \gray{} flux with respect to the optical one of less than 1 day.
Another remarkable result obtained by investigating all the available X-ray data accumulated during the time-span
2007--2010 is shown in Figure~\ref{fig:Xray_hysteresis}.
During the 18-months \agile{} campaign, 
\cite{Vercellone2010ApJ_P3} found a clear trend, in particular for fluxes 
above (1--2)$\times 10^{-11}$\,\ergcmsec\,.  
We can describe the harder-when-brighter trend in terms of a dominant contribution
of the external Compton (EC) off the disk seed photons, EC(Disk), over the synchrotron self-Compton
(SSC) component, probably due to an increase of the accretion rate. 
The constant X-ray photon index during the extreme \gray{} flares in 2009 and 2010 can be 
interpreted in terms of a balance of the SSC contribution with respect to the EC(Disk),
due to the possible increase of $\gamma_{\rm b}$ during the super-flares with respect to the
value during the less energetic \gray{} flares \cite{Vercellone2011:ApJL_3C454_nov2010}.
The net result is a roughly achromatic increase of the X-ray emission.

	\section{SHORT-TERM MONITORING}\label{sec:short}

\begin{figure}[!ht]
\centering
\includegraphics[width=80mm,angle=0]{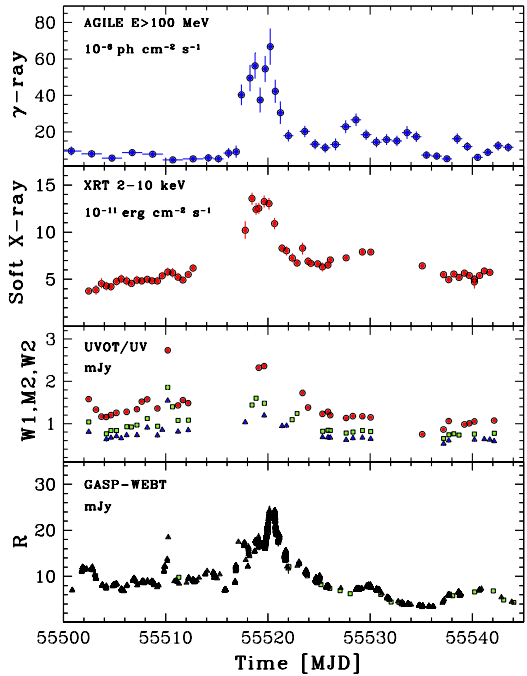}
\caption{From top to bottom: \gray{} ($E>100$\,MeV), X-ray (2--10\,keV),  UV ($w1$, $m2$, w2),
and optical ($R$) light-curves collected before, during, and after the 2010 November 20 (MJD~55520) 
\gray{} super-flare. The ``\gray-orphan'' optical-UV flare is visible at MJD~55510.} \label{fig:opt_flare}
\end{figure}

Detailed multi-wavelength campaigns were carried out during \gray{} flares in order to investigate the jet 
properties and to shed light on the radiation mechanisms responsible for the emission at the
different wavelengths.
\begin{figure}[!ht]
\centering
\includegraphics[width=80mm,angle=0]{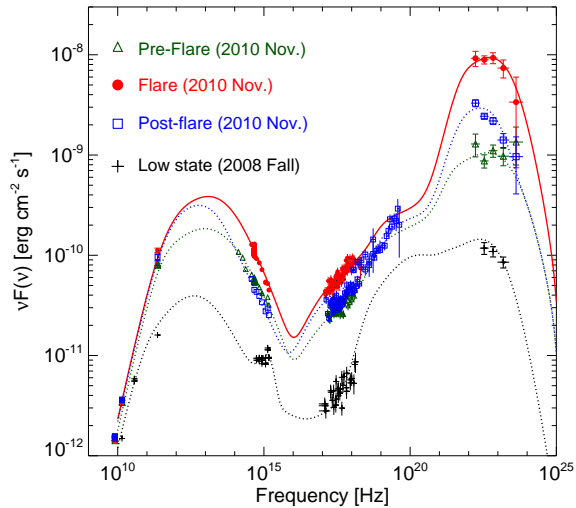}
\caption{
SEDs accumulated during the 2010 November flare (in colors, \cite{Vercellone2011:ApJL_3C454_nov2010}) 
compared with a SED accumulated during a particularly low \gray{} state in Fall 2008 (in black, \cite{Vercellone2010ApJ_P3}).
} \label{fig:seds}
\end{figure}
Figure~\ref{fig:seds} shows the spectral energy distributions (SEDs) during the 2010 November
super-flare (color points and lines) compared with the SED during a low \gray{} state in 2008 Fall
(black points and line). As noted in Section~\ref{sec:long}, the dynamic range in the \gray{}
energy band spans about two orders of magnitude, while the SSC peak dynamic range is
about one order of magnitude. The super-flare SEDs modeling have to take into account also
the flux variability at different wavelengths. About 10 days prior to this super-flare, we detected
a so-called ``\gray-orphan'' optical-UV flare, as shown in Figure~\ref{fig:opt_flare}.
A possible explanation of this complex behavior is as follows: 1) an energetic particle ignition causes the first 
optical flare at MJD~55510. 2) Subsequently, the blob moves away by 
$(c\,t\, \delta)/(1+z) \approx 3.4\times10^{17}$\,cm ($t=7$\,d), toward a region with a denser
external photon field in which a doubling in the optical flux can be followed by a stronger EC
counterpart, as observed during the \gray{} enhanced emission at MJD~55517. 3) Since the blob is moving 
in a region with enhanced density of external seed photons, the optical
and \gray{} flux variations have similar dynamic range (as observed at MJD~55520), until the blob
leaves this denser region. 4) Subsequently, as observed in the post-flare SED, the \gray{} emission 
decreases because of both the radiative cooling and the decrease of the external photon field due to the 
blob escaping the enhanced density region. In our modeling, the \gray{} dissipation region lies within
the broad-line region (BLR).

	\section{CONCLUSIONS}\label{sec:concl}

The extreme \gray{} variability (see e.g., \cite{Foschini2011:fsrq_variab}) of \source{} is still to be 
fully understood, and different models with respect to the one presented here can be invoked to explain
the observed SEDs (see e.g., \cite{Jorstad2010:3C454MW,Bonnoli2011:3c454}). 
An intriguing possibility to explain such an extreme behavior is to invoke the presence of a 
super-massive binary black-hole, as suggested in \cite{Qian2007:3C454_smbbh}. This
hypothesis is one of the possible challenges for future long-term projects.

\bigskip 
	\begin{acknowledgments}
We acknowledge financial contribution from the agreement ASI-INAF I/009/10/0.
The AGILE Mission is funded by the Italian Space Agency (ASI) with
scientific and programmatic participation by the Italian Institute
of Astrophysics (INAF) and the Italian Institute of Nuclear
Physics (INFN). We acknowledge ASI contract I/089/06/2.
\end{acknowledgments}

\bigskip 


\begin{thebibliography}{10}
\providecommand{\enquote}[1]{``#1''}
\expandafter\ifx\csname url\endcsname\relax
  \def\url#1{\texttt{#1}}\fi
\expandafter\ifx\csname urlprefix\endcsname\relax\def\urlprefix{URL }\fi

\bibitem{Hartman1992:3C454iauc}
R.~C. {Hartman}, et al., \emph{\iaucirc} \textbf{5477}, 2 (1992).

\bibitem{Hartman1993:3C454_EGRET}
R.~C. {Hartman}, et al., \emph{\apjl} \textbf{407}, L41--L44 (1993).

\bibitem{Aller1997:3C454_EGRET}
M.~F. {Aller}, et al., \enquote{{Radio to gamma Ray Observations of 3C 454.3:
  1993-1995},} in \emph{Proceedings of the Fourth Compton Symposium}, edited by
  C.~D. {Dermer}, M.~S. {Strickman}, and J.~D. {Kurfess}, 1997, vol. 410 of
  \emph{American Institute of Physics Conference Series}, p. 1423.

\bibitem{Giommi2006:3C454_Swift}
P.~{Giommi}, et al., \emph{\aap} \textbf{456}, 911--916 (2006).

\bibitem{vil06}
M.~{Villata}, et al., \emph{\aap} \textbf{453}, 817--822 (2006).

\bibitem{Pian2006:3C454_Integral}
E.~{Pian}, et al., \emph{\aap} \textbf{449}, L21--L25 (2006).

\bibitem{vil07}
M.~{Villata}, et al., \emph{\aap} \textbf{464}, L5--L9 (2007).

\bibitem{rai07}
C.~M. {Raiteri}, et al., \emph{\aap} \textbf{473}, 819--827 (2007).

\bibitem{rai08a}
C.~M. {Raiteri}, et al., \emph{\aap} \textbf{485}, L17--L20 (2008).

\bibitem{rai08b}
C.~M. {Raiteri}, et al., \emph{\aap} \textbf{491}, 755--766 (2008).

\bibitem{Vercellone2008:3C454_ApJ}
S.~{Vercellone}, et al., \emph{\apjl} \textbf{676}, L13--L16 (2008).

\bibitem{Vercellone2008:3c454:ApJ_P1}
S.~{Vercellone}, et al., \emph{\apj} \textbf{690}, 1018--1030 (2009).

\bibitem{Donnarumma2009ApJ_P2}
I.~{Donnarumma}, et al., \emph{\apj} \textbf{707}, 1115--1123 (2009).

\bibitem{Vercellone2010ApJ_P3}
S.~{Vercellone}, et al., \emph{\apj} \textbf{712}, 405--420 (2010).

\bibitem{Pacciani2010ApJ_3c454}
L.~{Pacciani}, et al., \emph{\apjl} \textbf{716}, L170--L175 (2010).

\bibitem{Striani2010ApJ_3c454}
E.~{Striani}, et al., \emph{\apj} \textbf{718}, 455--459 (2010).

\bibitem{Vercellone2011:ApJL_3C454_nov2010}
S.~{Vercellone}, et al., \emph{\apjl} \textbf{736}, L38+ (2011).

\bibitem{Bulgarelli2009:ASPC_DataAnalysis}
A.~{Bulgarelli}, M.~{Trifoglio}, F.~{Gianotti}, G.~{Di Cocco}, and M.~{Tavani},
  \enquote{{AGILE-GRID Automated Web-based Analysis System for Fast Detection
  of Gamma-ray Transients},} in \emph{Astronomical Data Analysis Software and
  Systems XVIII}, edited by {D.~A.~Bohlender, D.~Durand, \& P.~Dowler}, 2009,
  vol. 411 of \emph{Astronomical Society of the Pacific Conference Series}, pp.
  362--+.

\bibitem{rai11}
C.~M. {Raiteri}, et al., \emph{\aap} \textbf{534}, A87+ (2011).

\bibitem{Foschini2011:fsrq_variab}
L.~{Foschini}, G.~{Ghisellini}, F.~{Tavecchio}, G.~{Bonnoli}, and
  A.~{Stamerra}, \emph{\aap} \textbf{530}, A77+ (2011).

\bibitem{Jorstad2010:3C454MW}
S.~G. {Jorstad}, et al., \emph{\apj} \textbf{715}, 362--384 (2010).

\bibitem{Bonnoli2011:3c454}
G.~{Bonnoli}, G.~{Ghisellini}, L.~{Foschini}, F.~{Tavecchio}, and
  G.~{Ghirlanda}, \emph{\mnras} \textbf{410}, 368--380 (2011).

\bibitem{Qian2007:3C454_smbbh}
S.-J. {Qian}, et al., \emph{\cjaa} \textbf{7}, 364--374 (2007).

\end{thebibliography}

\end{document}